\documentclass{article}

\usepackage{arxiv}

\usepackage[utf8]{inputenc} 
\usepackage[T1]{fontenc}    
\usepackage{hyperref}       
\usepackage{url}            
\usepackage{booktabs}       
\usepackage{amsfonts}       
\usepackage{nicefrac}       
\usepackage{microtype}      
\usepackage{lipsum}		
\usepackage{graphicx}
\usepackage{natbib}
\usepackage{doi}
\usepackage{caption}
\usepackage{subcaption}

\usepackage{amsmath, multicol, multirow, booktabs, microtype, balance}
\usepackage{commath}
\usepackage{xcolor}

    \title{Convolutional Nets for Diabetic Retinopathy Screening in Bangladeshi Patients}

\author{ Ayaan Haque\thanks{Corresponding Author} \\
	Saratoga High School \\
	Saratoga, CA, USA\\
	BRAC University JPG School of Public Health \\ 
	Dhaka, Bangladesh \\
	\texttt{ayaanzhaque@gmail.com} \\
	\And
	Ipsita Sutradhar \\
	BRAC University JPG School of Public Health \\ 
	Dhaka, Bangladesh \\
	\texttt{ipsita.sutradhar@bracu.ac.bd} \\
	\AND
	Mahziba Rahman \\
	BRAC University JPG School of Public Health \\ 
	Dhaka, Bangladesh \\
	\texttt{mahziba@gmail.com} \\
	\And
	Mehedi Hasan \\
	BRAC University JPG School of Public Health \\ 
	Dhaka, Bangladesh \\
	\texttt{mehedi.hasan@bracu.ac.bd} \\
	\And
	Malabika Sarker \\
	BRAC University JPG School of Public Health \\ 
	Dhaka, Bangladesh \\
	\texttt{malabika@bracu.ac.bd} \\
}

\date{}
\renewcommand{\shorttitle}{CNNs for DR Screening in Bangladeshi Patients}

\begin{document}
\maketitle

\begin{abstract}
	Diabetes is one of the most prevalent chronic diseases in Bangladesh, and as a result, Diabetic Retinopathy (DR) is widespread in the population. DR, an eye illness caused by diabetes, can lead to blindness if it is not identified and treated in its early stages. Unfortunately, diagnosis of DR requires medically trained professionals, but Bangladesh has limited specialists in comparison to its population. Moreover, the screening process is often expensive, prohibiting many from receiving timely and proper diagnosis. To address the problem, we introduce a deep learning algorithm which screens for different stages of DR. We use a state-of-the-art CNN architecture to diagnose patients based on retinal fundus imagery. This paper is an \textit{experimental evaluation} of the algorithm we developed for DR diagnosis and screening specifically for Bangladeshi patients. We perform this validation study using separate pools of retinal image data of real patients from a hospital and field studies in Bangladesh. Our results show that the algorithm is effective at screening Bangladeshi eyes even when trained on a public dataset which is out of domain, and can accurately determine the stage of DR as well, achieving an overall accuracy of 92.27\% and 93.02\% on two validation sets of Bangladeshi eyes. The results confirm the ability of the algorithm to be used in real clinical settings and applications due to its high accuracy and classwise metrics. Our algorithm is implemented in the application \textit{Drishti}, which is used to screen for DR in patients living in rural areas in Bangladesh, where access to professional screening is limited. \footnote{Our code is available at \href{https://github.com/Drishti-BD/Drishti-CNN}{https://github.com/Drishti-BD/Drishti-CNN}}
\end{abstract}

\keywords{Diabetic Retinopathy \and Deep Learning \and CNN \and Classification \and Experimental Study \and Clinical Data}

\section{Introduction}
\label{sec:intro}

Diabetes in Bangladesh is a high-risk disease that affects millions. Specifically, 8.4 million adults in Bangladesh lived with diabetes in 2019, and the number is projected to double by 2045 to 15.0 million. Moreover, 3.8 million people lived with prediabetes. The prevalence of diabetes in the population has steadily increased in terms of percentage of population (\cite{islam2021prevalence}). Diabetic Retinopathy (DR) is a complication of diabetes that causes degeneration of eyesight due to damaged blood vessels in the retina. This can be due to high blood sugar which blocks small blood vessels that supply blood to the retina. Early diabetic retinopathy may not have any symptoms, making it challenges to detect early by patients themselves. Symptoms of diabetic retinopathy include spots in vision, blurred vision, fluctuating vision, and/or dark areas in vision. 

If undetected, DR can progress to cause blindness. It is a leading cause of blindness across the world (\cite{porta2002diabetic}). Thus, early screening processes are required in order to prevent DR from causing blindness. In Bangladesh, there is a shortage of DR specialists, meaning many patients who have DR do not have access to proper screening. Moreover, most specialists are located in Dhaka, the capital, and therefore patients with diabetes outside of Dhaka of Bangladesh are deterred from visiting specialists. Most importantly, screenings performed by eye specialists are expensive, and when these screenings must be done in a routine manner, many patients simply do not have the finances to be screened. Thus, without affordable and accessible screening, Bangladeshi diabetes patients are developing DR at high rates. Our goal is to deploy free and accurate AI algorithms in rural areas to provide care for disadvantaged and vulnerable populations.

Fundus imaging is an effective method of diagnosing DR (\cite{muqit2019trends}). Deep learning for medical imaging analysis has received great interest in recent decades, and using such systems can be even more effective at diagnosing DR. DR classification follows a general standardized scale (further described in Section \ref{sec:data-prep}), with two main categories: nonproliferative and proliferative. If DR is detected while it is nonproliferative, it can be properly controlled (\cite{safi2018early}). However, this requires early intervention predicated on timely screenings.

Many deep learning CNNs have been proposed to perform binary classification of DR (\cite{gargeya2017automated, lam2018automated, tymchenko2020deep}). There have also been other methods which classify DR into separate stages (\cite{arcadu2019deep, dutta2018classification, islam2018deep}), like our work. All of these methods use fundus imagery. Previously, most machine learning algorithms used for diagnosis did not employ CNN architectures but rather used standard ML algorithms such as Gaussian Mixture Models (GMMs), K-Nearest Neighbors (kNNs), and Support Vector Machines (SVMs). Newer methods all utilize popular CNN architecture such as the VGG (\cite{simonyan2015deep}), ResNet (\cite{he2016deep}), InceptionNet (\cite{szegedy2015going}) and their variants, often pre-trained on ImageNet. Some methods have even employed ensemble or cascaded networks of entirely CNNs or using segmentation architectures such as U-Nets followed by CNNs. Regarding training procedures, these methods often utilize data augmentation and other generalization techniques. In our algorithm, we use many techniques and strategies proposed in the literature presented.

This paper is a clinical validation of our algorithm to screen for stages of DR. The algorithm and methods itself is not individually innovative, and this is not our goal. Rather, the application and integration are the focus of this paper. This paper presents the effectiveness of our AI algorithm in detecting DR in Bangladeshi patients. Due to the domain generalization problem in medical imaging, a specific study on Bangladeshi fundus images is required to show our algorithm can be properly deployed in our target regions.

\section{Methods}
\label{sec:methods}

\subsection{Model Architecture}

\begin{figure}
    \centering
    \includegraphics[scale=0.3]{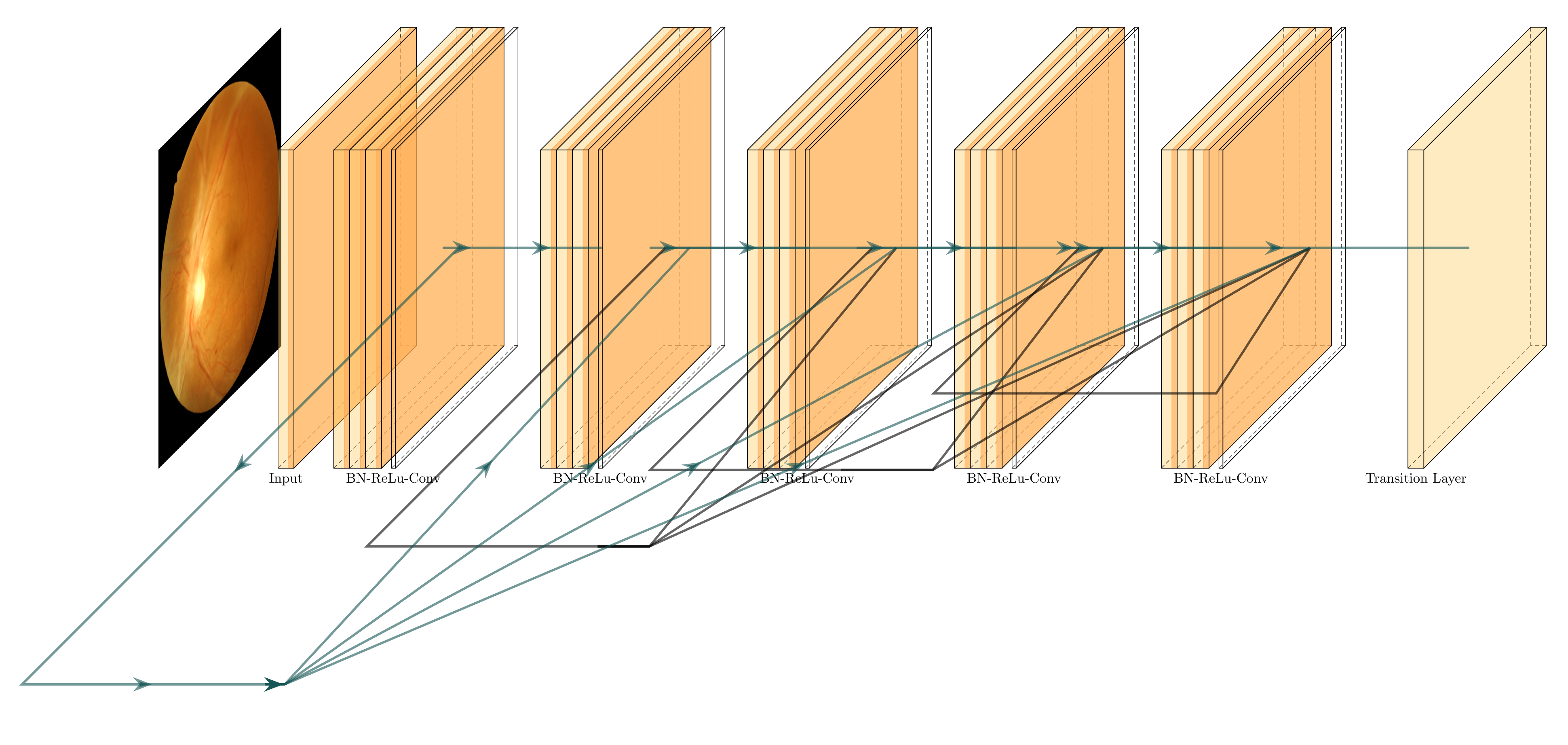}
    \caption{Diagram of a DenseNet-121 convolutional block with our retinal scans.
    }
    \label{Fig:densenet-retina}
\end{figure}

\begin{figure}
    \centering
    \includegraphics[scale=0.75]{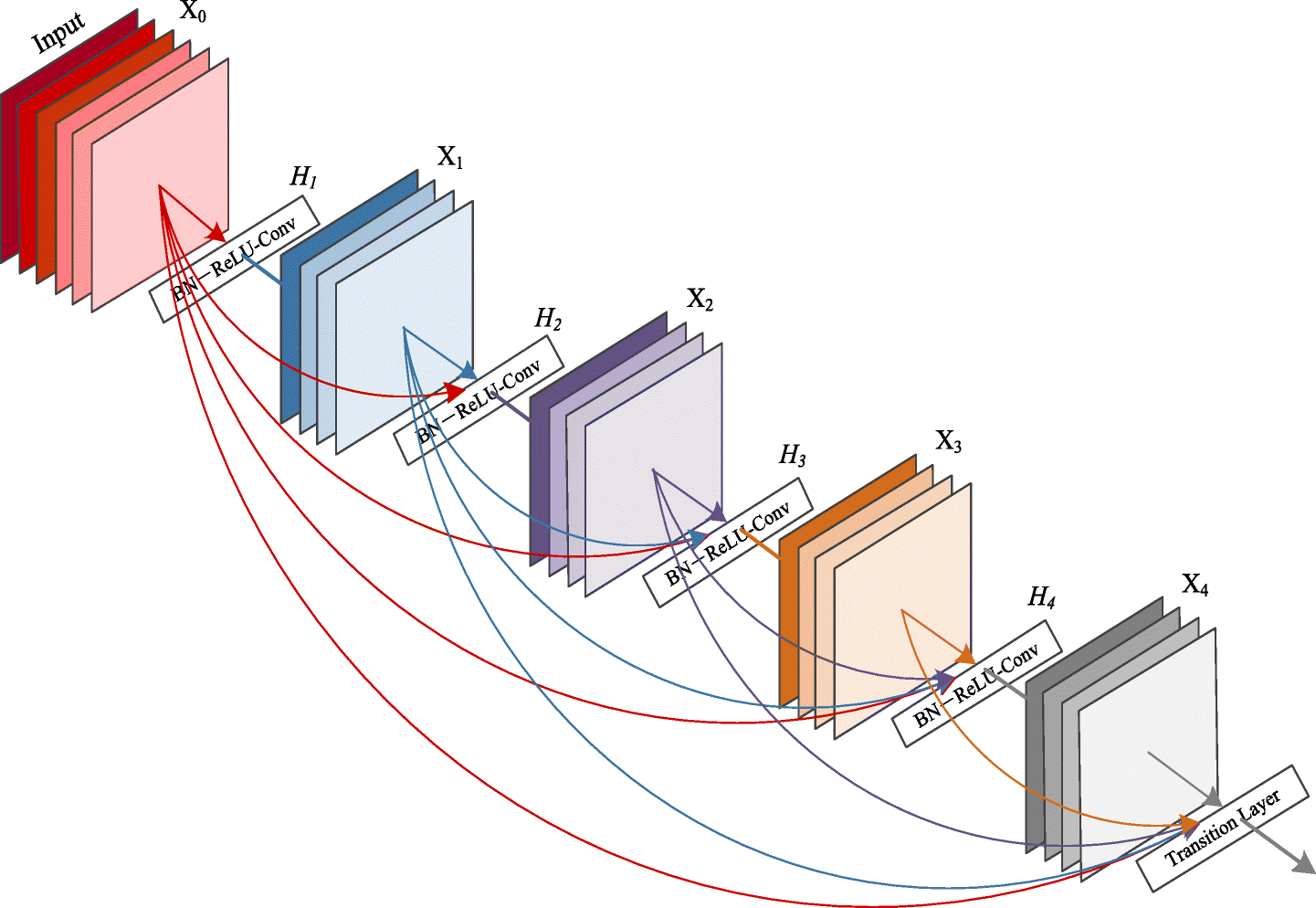}
    \caption{Diagram of the DenseNet-121 architecture (\cite{huang2017densely}).
    }
    \label{Fig:densenet}
\end{figure}

We use a DenseNet-121 CNN (\cite{huang2017densely}) as the base network. The DenseNet is a state-of-the-art classification architecture that has been used in many other medical imaging classification methods (\cite{xu2018improved, gottapu2018densenet, zhou2018weakly}). Importantly, we do not use a pre-trained version of the DenseNet-121 and rather train entirely with randomized parameters, as we found this increased the efficiency of training and deployment. The architecture was chosen after initial experimentation with other architectures, such as ResNet-50 or VGG-19. The base DenseNet with the output layer removed is used as the primary architecture of our network. A 2D Global Average Pooling is then added after the DenseNet module, followed by a Dropout layer with a dropout rate of 0.5. For the output layer, a dense layer with 5 output nodes and a sigmoid activation function are used. A diagram provided by the authors of DenseNet is shown in Figure \ref{Fig:densenet}. Figure \ref{Fig:densenet-retina} shows a DenseNet-121 convolutional block with our retinal images. This results in a network with a total of 7,042,629 parameters. 5 nodes are used for each of the separate classes used to classify DR. Further details on the classification scale is provided in Section \ref{sec:data-prep}. A detailed architecture is described in Table \ref{tab:architecture}.

\begin{table}
	\caption{Our DenseNet-121 architectural details. M represents the minibatch size.}
	\centering
	\begin{tabular}{llll}
		\toprule
        Layer & \phantom{a} & Input Size & Output Size\\
        \midrule
        DenseNet-121 Base && M $\times$ 224 $\times$ 224 $\times$ 3 & M $\times$ 7 $\times$ 7 $\times$ 1024\\
        \midrule
        Global Average Pooling && M $\times$ 7 $\times$ 7 $\times$ 1024 & M $\times$ 1024\\
        \midrule
        Dropout && M $\times$ 1024 & M $\times$ 1024\\
        \midrule
        Final Dense Layer && M $\times$ 1024 & M $\times$ 5\\
        \bottomrule
	\end{tabular}
	\label{tab:architecture}
\end{table}

The CNN is optimized using a conventional cross-entropy loss and the Adam variant of the SGD optimizer (\cite{kingma2015adam}). Regarding hyperparameters, we used a learning rate of 0.00005, a batch size of 32, and train for 15 epochs. The model was trained a total of 10 separate times, and the model with the best performance on the test set was used to perform the evaluation on Bangladeshi patients. Our code was written in the newest versions of Tensorflow and python and were trained using an NVIDIA K80 GPU.

\subsection{Data Preparation}
\label{sec:data-prep}

Our algorithm is trained on a public dataset of retinal images and validated on two datasets of retinal images collected from hospitals and field studies in Bangladesh.

Each image is graded on a 5-point scale as according to the International Classification of Diabetic Retinopathy (\cite{wong2018guidelines}). The grading is as follows:

\begin{itemize}
    \item 0: No Apparent DR
    \item 1: Mild Nonproliferative DR
    \item 2: Moderate Nonproliferative DR
    \item 3: Severe Nonproliferative DR
    \item 4: Proliferative DR
\end{itemize}

\begin{figure}
    \centering
    \begin{subfigure}{0.15\textwidth}
        \includegraphics[width=\hsize]{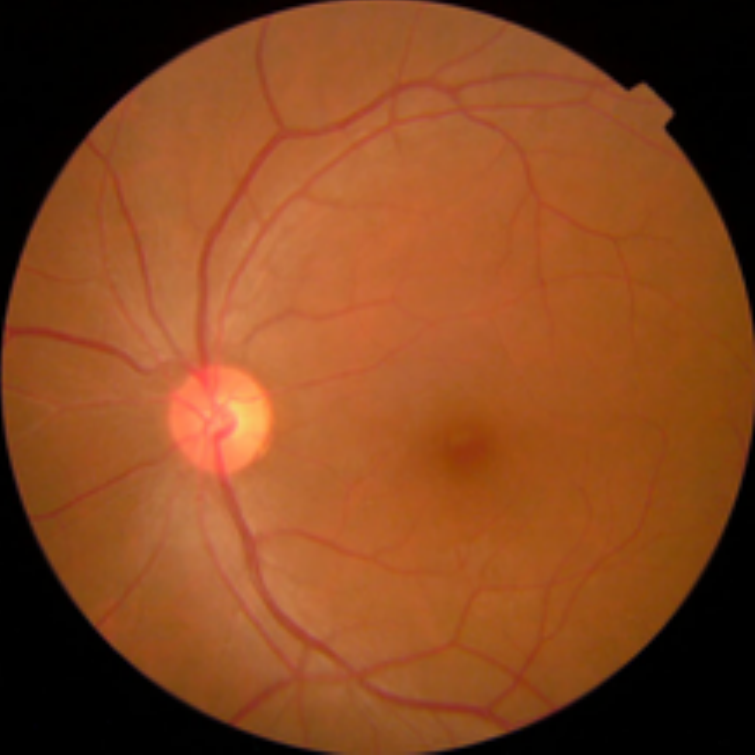}
        \caption{No DR}
        \label{fig:0}
    \end{subfigure}
    \begin{subfigure}{0.15\textwidth}
    \includegraphics[width=\hsize]{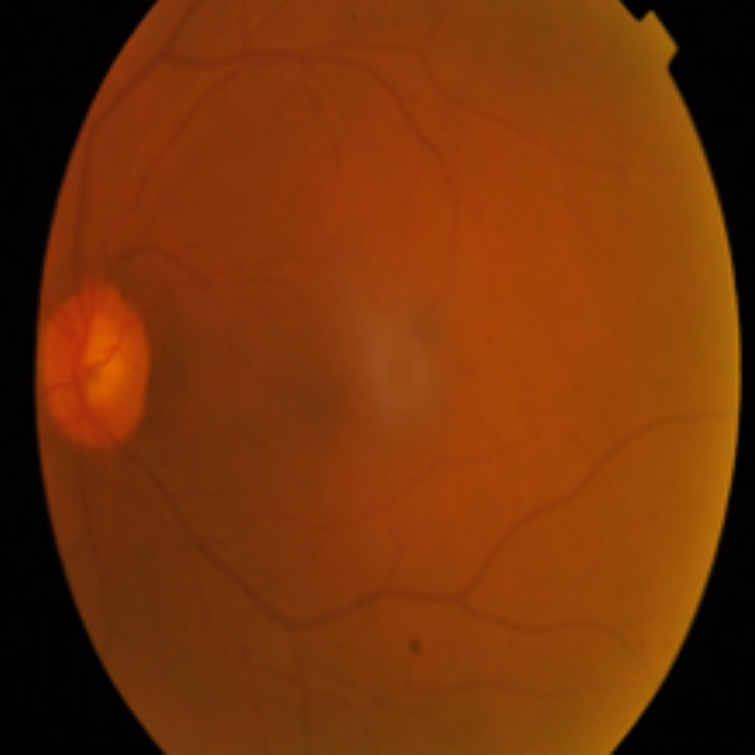}
    \caption{Mild DR}
    \label{fig:1}
\end{subfigure}
    \begin{subfigure}{0.15\textwidth}
    \includegraphics[width=\hsize]{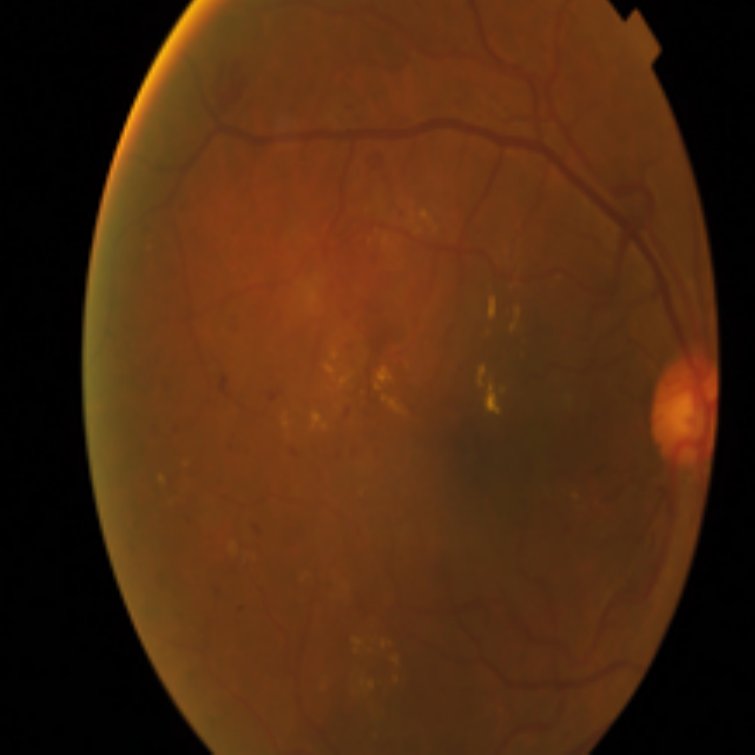}
    \caption{Moderate DR}
    \label{fig:2}
\end{subfigure}
    \begin{subfigure}{0.15\textwidth}
    \includegraphics[width=\hsize]{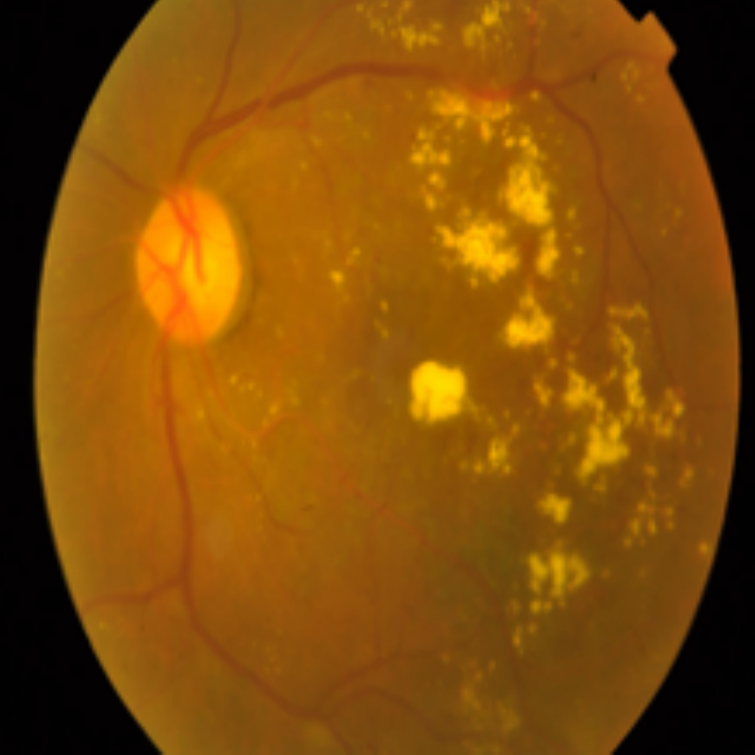}
    \caption{Severe DR}
    \label{fig:3}
\end{subfigure}
    \begin{subfigure}{0.15\textwidth}
    \includegraphics[width=\hsize]{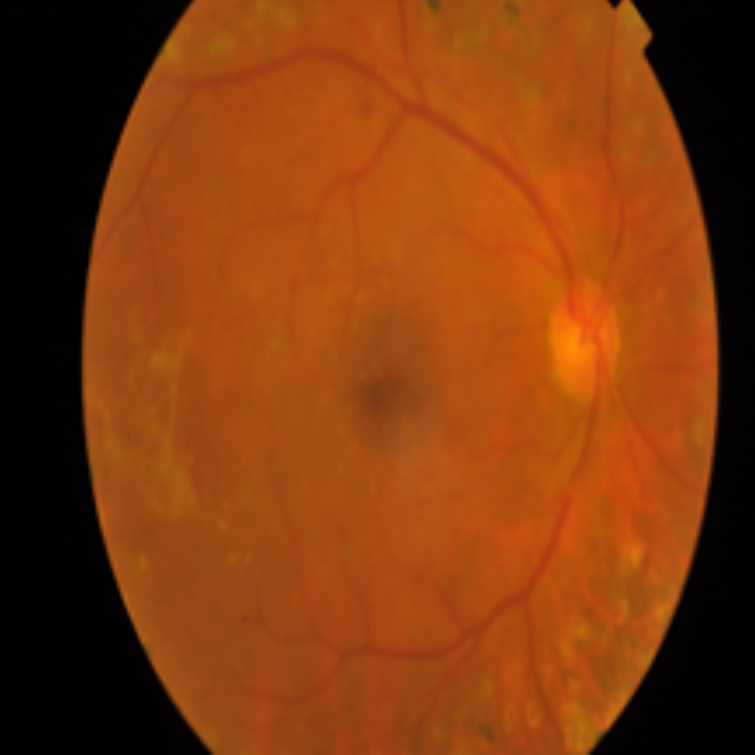}
    \caption{Proliferative DR}
    \label{fig:4}
\end{subfigure}
    \caption{Examples of fundus images from each DR stage from the APTOS training dataset.}
    \label{fig:aptos-examples}
\end{figure}

The labeling scheme for our training and validation data follows this scale. For example, a label of 0 corresponds to no apparent DR, and a label of 4 corresponds to proliferative DR. This makes the task a multi-class classification task. Examples of an image from each class in the training set is shown in Figure \ref{fig:aptos-examples}.

To pre-process the images, we first normalized the pixel intensities of the fundus images. Each image was then resized into a 224 $\times$ 224 $\times$ 3 image, as the DenseNet-121 requires the input to be in this specific shape. All images remained in the 3-channel RGB format. We used data augmentation to increase the performance of the CNN. These procedures include random zooming, horizontal flipping, and vertical flipping. For random zoom was applied with a probability of 0.15 and both horizontal and vertical flipping were applied with a probability of 0.5. 

\section{Data and Clinical Details}
\label{sec:data}

\subsection{Training}

\begin{figure}
    \centering
    \begin{tabular}{ll}
    \includegraphics[scale=0.35]{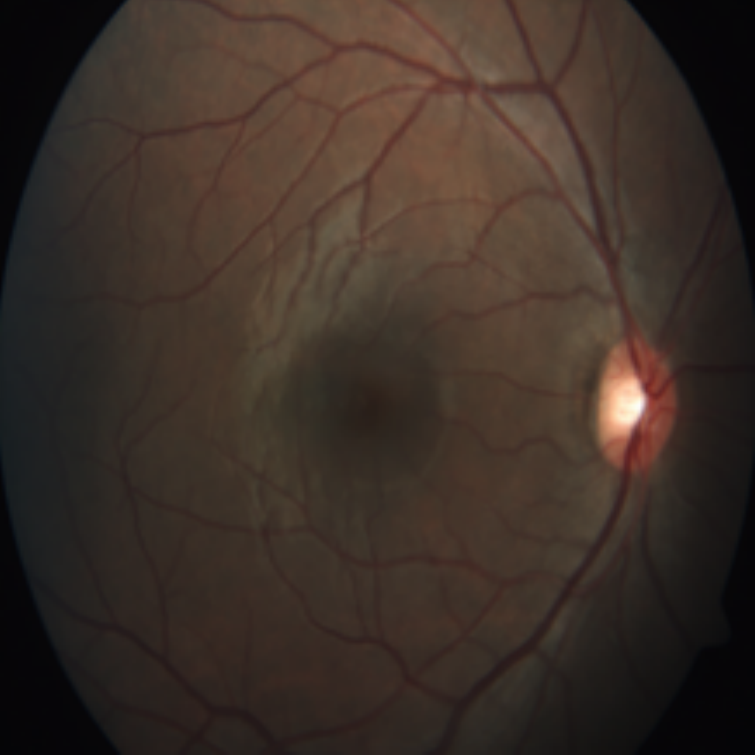}
    
    \includegraphics[scale=0.35]{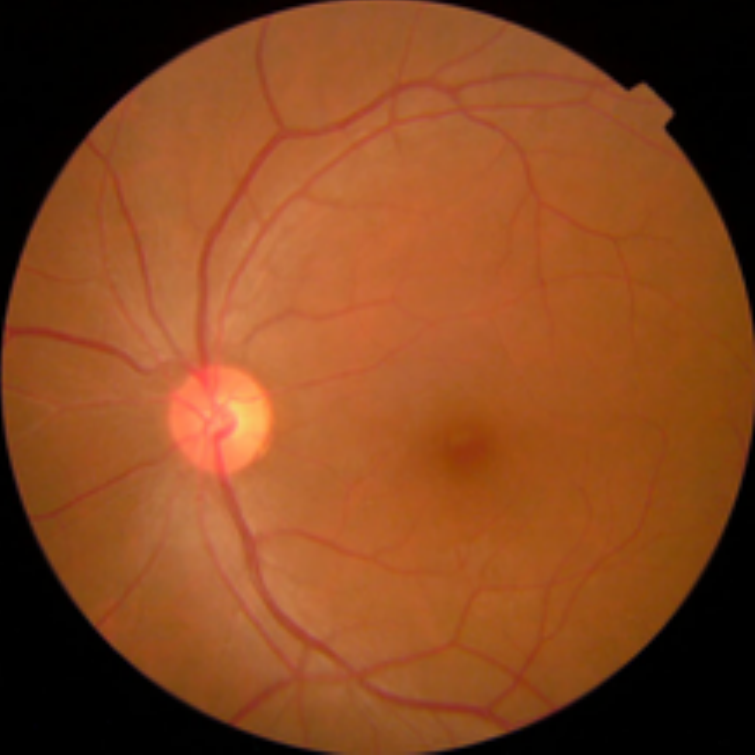}
    \\
    \includegraphics[scale=0.35]{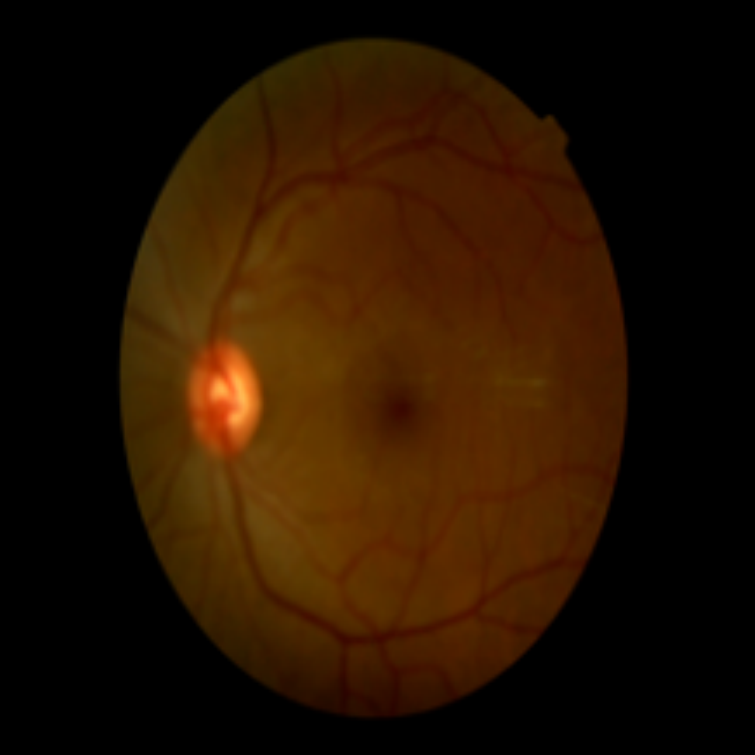}
    
    \includegraphics[scale=0.35]{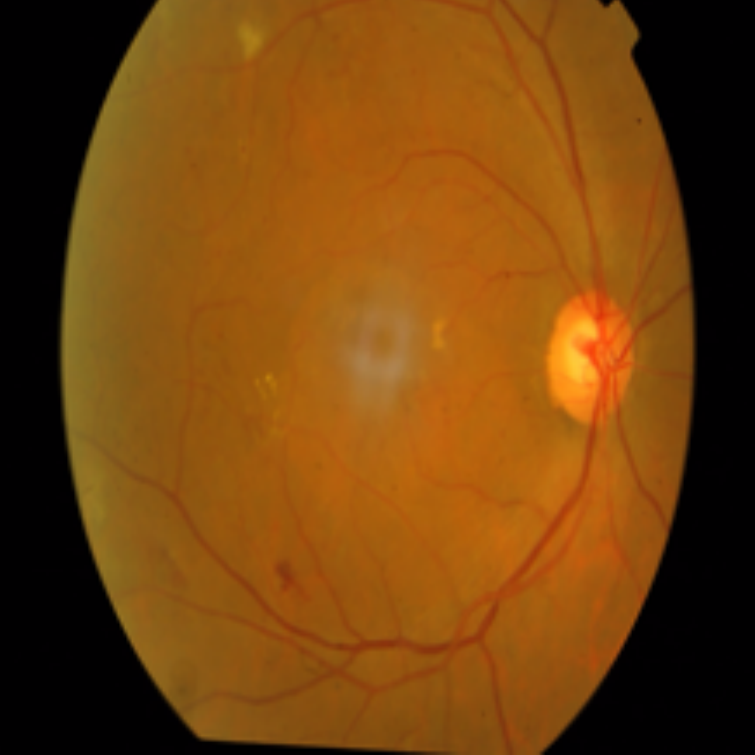}
    \end{tabular}
    \caption{Examples of drastically different training images from the APTOS dataset. A visible disparity in camera angle, coloring, lighting, and various other factors is noticeable in the images, making the classification task substantially more difficult.
    }
    \label{Fig:aptos-examples}
\end{figure}

For training the CNN, we used the open-source APTOS 2019 Blindness Detection fundus image dataset (\cite{aptos}). The dataset is labeled using the same grading scale as described in Section \ref{sec:data-prep}. Since it would be infeasible to collect enough data from our field studies to build an effective training set, we used a public dataset. However, we were not able to validate on this dataset as the provided images from the testing set do not have provided labels or diagnoses. The APTOS dataset is a collection of retinal fundus images from rural areas in India. Therefore, training on this dataset and deploying it in the cross-domain setting of rural Bangladesh is reasonable due to geographical proximity and demographical similarity. The details of the APTOS dataset are found in Table \ref{tab:aptos-data}. The images are collected from a variety of clinics and health centers, meaning the camera angle, lighting, coloring, and artifacts seen in the images will vary and increase the difficulty of the classification task. Examples of fundus images from the APTOS dataset are shown in Figure \ref{Fig:aptos-examples}. 

\begin{table}[h]
	\caption{Details of the APTOS dataset.}
	\centering
	\begin{tabular}{lllll}
		\toprule
        Dataset & \phantom{a} & Total & Train & Test\\
        \midrule
        APTOS 2019 && 5,590 & 3,662 & 1,928\\
        \bottomrule
	\end{tabular}
	\label{tab:aptos-data}
\end{table}

\subsection{Validation}

To validate the performance of the algorithm in the target domain of Bangladeshi patients, we utilized two separate datasets. We first performed a smaller validation study with the Bangladesh Eye Hospital. The associate professor and lead diabetic retinopathy specialist, Dr. Sabrina Rahmatullah, provided us with 43 images from 23 patients to validate with. There was an even distribution of left and right eyes. These retinal images were acquired with a Topcon angiogram machine. The images were first independently graded and diagnosed by the specialist. Then the images were supplied to us and we performed inference with the algorithm on the images after proper pre-processing and data preparation. These fundus images were collected from patients at the hospital using their funduscope. In addition to the grading, each image was also given a confidence rating so that more proper evaluation could be made if the algorithm prediction did not match. All patient information from the fundus images was kept confidential and not provided to us during the study, as they are not needed for screening purposes.

The second set of images was provided by BRAC University School of Public Health. The set of images contains 206 images from 103 patients. The images were graded by researcher and trained specialist Dr. Mahziba Rahman, and the inference from our algorithm was performed simultaneously. These images were collected using a handheld camera, meaning the images are acquired in a similar format as our potential application setting. The labels and the predictions were than compared and not modified after. Each image was given a confidence rating by the specialist, which was used in scenarios when the diagnosis from the algorithm and specialist did not match.

All of these images were pre-processed separately than the training images. Each image used for validation was manually cropped so there were no distortions in the image and so the retina was in the center of the image. They were resized into square dimensions and then were downsized into a 224 $\times$ 224 $\times$ 3 format. These images from either validation set cannot be shown or publicly provided due to patient confidentiality. Patient information for these scans were also not provided to us during this study. A summary of the validation sets is shown in Table \ref{tab:validation-data}, including number of scans, number of patients, and the timeframe of collection.

\begin{table}[h]
	\caption{Details of the validation sets.}
	\centering
	\begin{tabular}{lcccl}
		\toprule
        Dataset & \phantom{a} & Total Images & Patients & Timeframe of Collection\\
        \midrule
        Bangladesh Eye Hospital && 43 & 23 & August - October 2020\\
        \midrule
        BRAC University && 206 & 103 & ---\\
        \bottomrule
	\end{tabular}
	\label{tab:validation-data}
\end{table}

\section{Experimental Results}
\label{sec:results}

Table \ref{sec:results} shows the quantitative results of the CNN on both validation sets. The accuracy column represents the overall accuracy of our method at classifying the correct stage, not just a binary result. Therefore, we are able to achieve 92.27\% and 93.02\% accuracy at correctly predicting the exact stage of DR, not just whether or not it is DR positive. This means the CNN is very accurate at correctly identifying DR in patients through fundus imagery. To calculate binary accuracy, sensitivity and specificity, we convert the labels and predictions into a binary class problem of DR positive and DR negative. DR negative includes samples that are classified as no DR, and DR positive includes samples that are classified as mild, moderate, severe, and proliferative. For binary accuracy, we achieve even higher accuracies in this metric.  For medical purposes, we achieve high sensitivity and specificity metrics on both datasets, as they are both $>$ 90\%.

\begin{table}[h]
	\caption{Overall Accuracy and metrics for our performance on the two validation sets of Bangladeshi Eyes.}
	\centering
	\begin{tabular}{ccc}
		\toprule
        & Bangladesh Eye Hospital & BRAC University\\
        \midrule
        Overall Accuracy & 93.02 & 92.27\\
        \midrule
        Binary Accuracy & 100.00 & 93.23\\
        \midrule
        Sensitivity & 100.00 & 94.21\\
        \midrule
        Specificity & 100.00 & 92.83\\
        \bottomrule
	\end{tabular}
	\label{tab:results}
\end{table}

Of the 206 images from the field study, we were able to correctly diagnose 190/206 when compared to Dr. Rahman. Of the 16 images that were incorrectly diagnosed, 5 scans were within one grade difference (e.g. real label is 1 and algorithm predicts 2). Moreover, of the 40/43 images our algorithm correctly diagnosed according to Dr. Rahmatullah, all of the incorrectly predicted images were within a single grade difference, furthering confirming the strong performance of our algorithm on Bangladeshi eyes. For the Eye Hospital validation set, we were able to determine whether or not the patient had DR or not for every patient, which is a binary accuracy of 100\%.

Since the results on the training dataset are not the focus on this paper, we do not include thorough evaluation in this paper. However, it should be noted that our CNN achieves a 96.60\% accuracy of predicting the correct stage of DR on the APTOS training set, which is an impressive mark. However, this is an in-domain evaluation and not in our target domain, so this result should not be focused on.

\begin{figure}[h]
    \centering
    \includegraphics[scale=0.5]{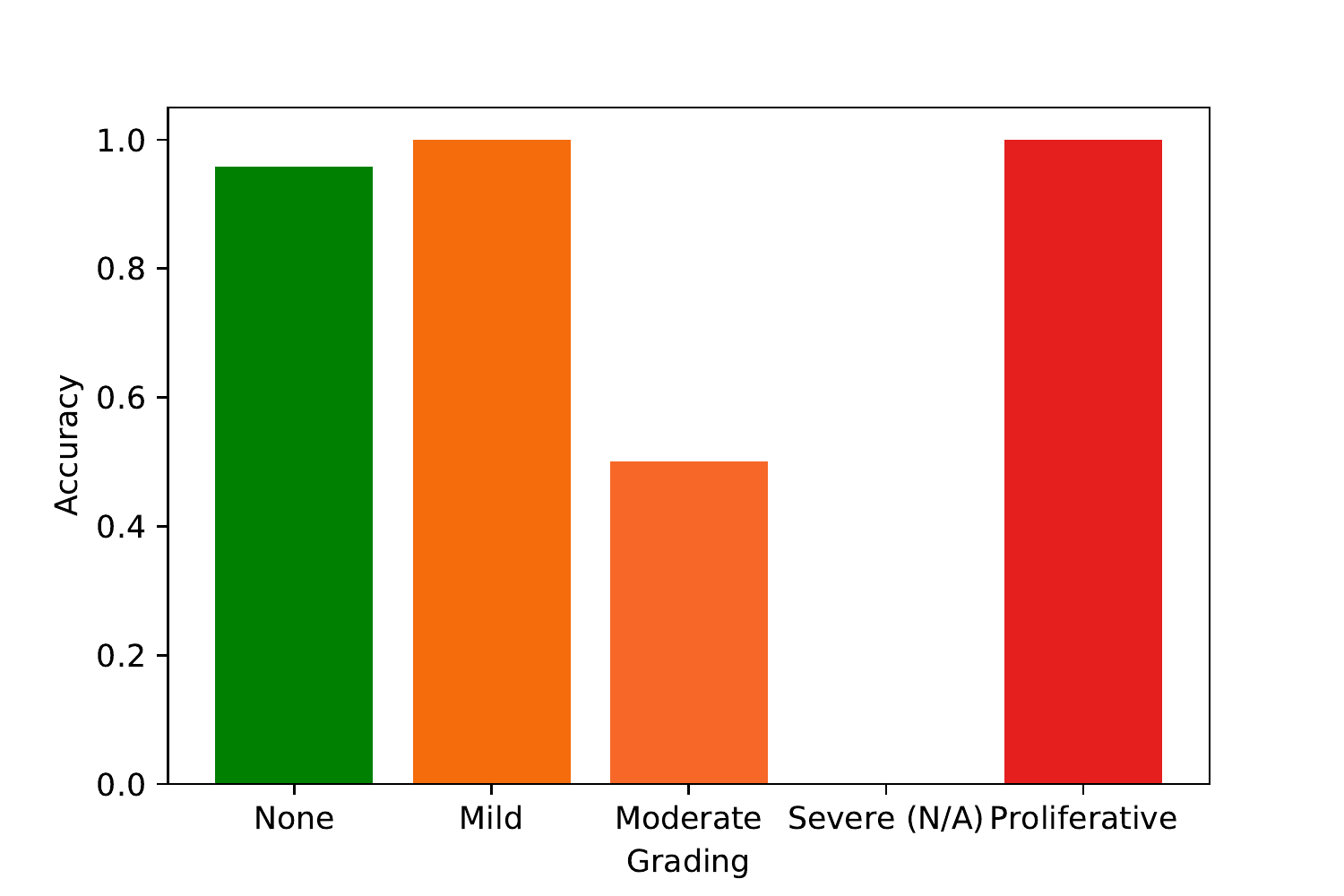}
    \includegraphics[scale=0.5]{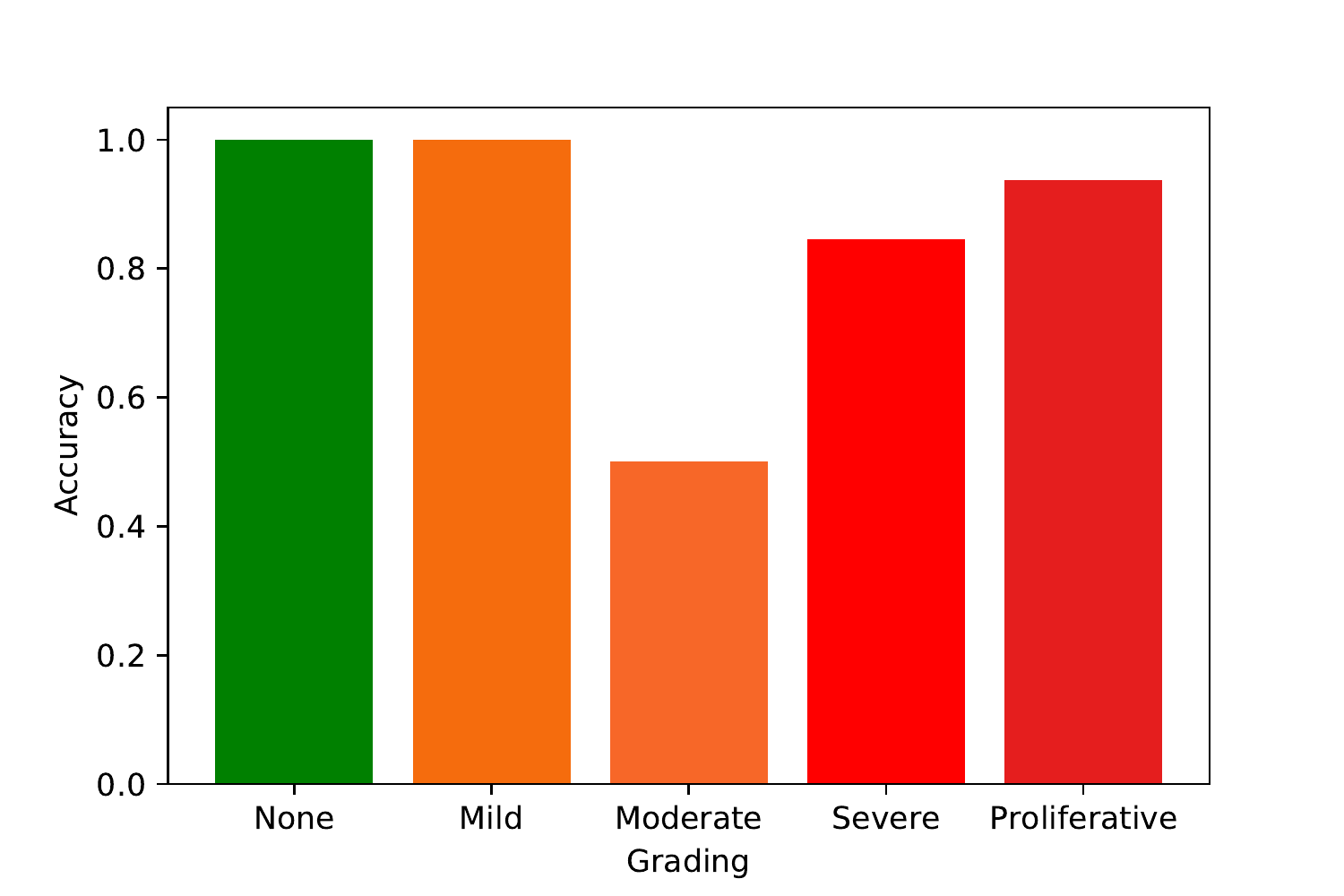}
    \caption{Accuracy of the CNN for each level of grading. The left figure is on the BRAC University dataset and the right is on the Bangladesh Eye Hospital dataset. Note that for the BRAC University dataset, none of the 206 images were graded as Stage 4, Severe DR).
    }
    \label{Fig:classwise-acc}
\end{figure}

The classwise accuracy of our network is shown in Figure \ref{Fig:classwise-acc}. As shown, our CNN is very accurate at classifying no DR. While the number of samples for mild and proliferative cases is low, we are able to classify them correctly every single time. While the accuracy for mild cases is relatively lower, generally the CNN's classification is within one level of grading, which is an adequate result when deployed in real settings.

\begin{figure}[h]
    \centering
    \includegraphics[scale=0.5]{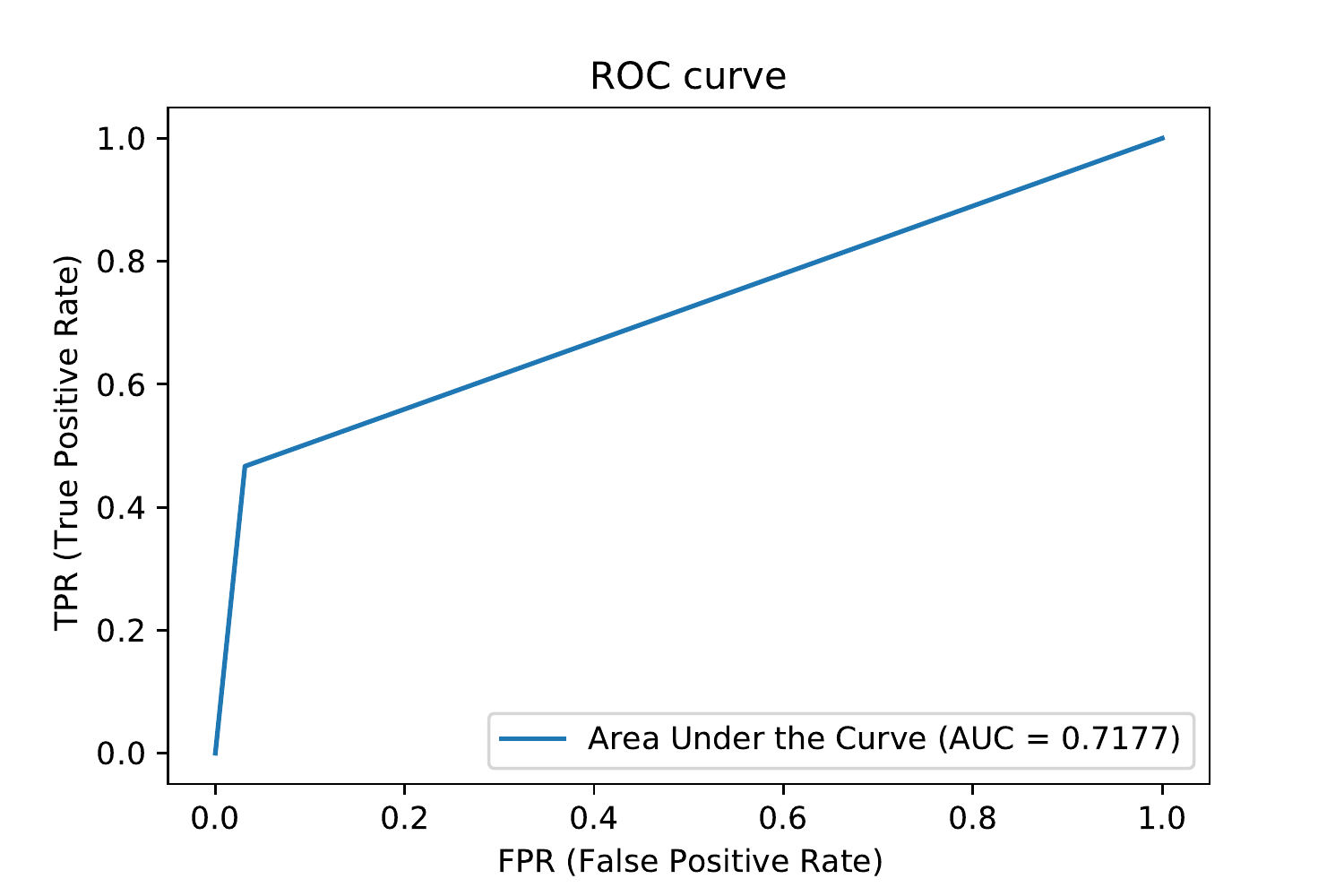}
    \includegraphics[scale=0.5]{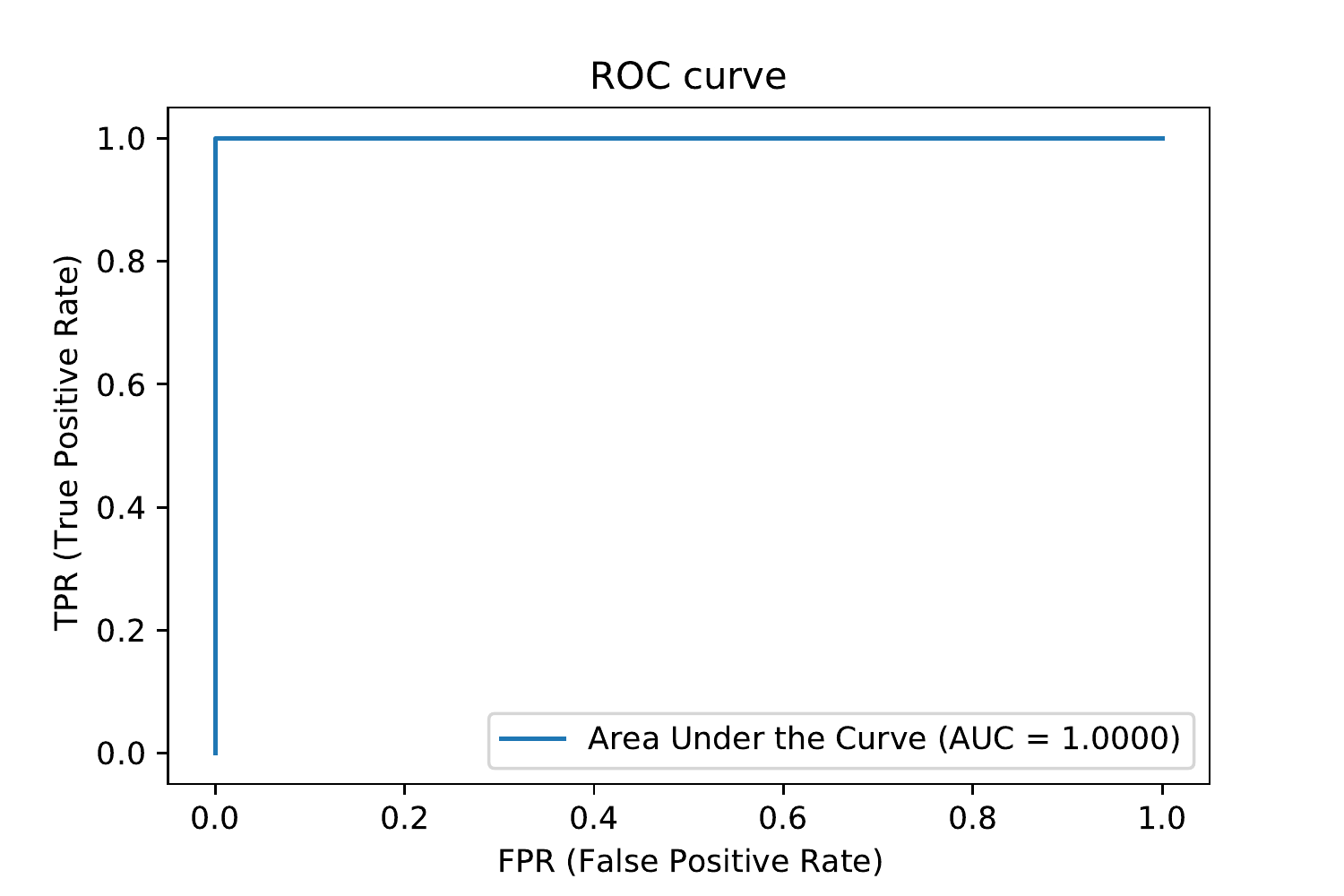}
    \caption{ROC curves for the binary classification of DR. The left figure is on the BRAC University dataset and the right is on the Bangladesh Eye Hospital dataset. All categories of DR are combined into a single class of DR. Our CNN is shown to have relatively strong performance on binary classwise accuracy.
    }
    \label{Fig:roc-curve}
\end{figure}

Figure \ref{Fig:roc-curve} shows an ROC curve for the binary prediction of DR positive and DR negative. The Area Under Curve value is relatively high and we achieve strongly desired true positive and false positive rates. This proves the efficacy of our AI algorithm to be used in real clinical settings. The results shown are impressive and can be very effective as initial screening procedures in areas where proper screening is unavailable.

\section{Conclusions and Discussion}
\label{sec:conclusions}

In this paper, we present an experimental validation of the introduced CNN for Diabetic Retinopathy screening on Bangladeshi eyes. We show that our algorithm can accurately screen for different stages of DR in real patients even when the data is collected in non-uniform methods. While our algorithm is trained on a dataset from a separate domain, using generalization techniques, we can effectively screen Bangladeshi patients and match the diagnostic performance of trained specialists who we collaborated with, which is a difficult task. The results on hundreds of images is promising for the future of our application, especially because the BRAC University images were collected with retinal cameras. Our future work for the algorithm itself will investigate using more deep learning techniques such as semi-supervised learning.

This algorithm is implemented in the app \textit{Drishti}, an organization which works with NGOs to provide free and early access to patients in rural areas without access to professional screening and treatment. Our algorithm aims to be deployed initially in pilot clinics and provide patients with routine check-ups and screenings. We make it evidently clear that we are not a 100\% accurate diagnosis and do not claim to do so ever. We are rather an initial screening process that allows patients who otherwise are not receiving any feedback on their eyes and their conditions to be more informed about their health. 

Our algorithm will be integrated with smartphone-based retinal cameras so that technicians, without the assistance of professional specialists, can operate the system and provide diagnosis. We are currently using the D-EYE smartphone retinal camera studied and introduced by \cite{russo2016comparison} and \cite{mamtora2018smart}. The fundus images taken by the low-cost system can be easily fed into our CNN, which is hosted on \textit{Drishti}'s web server and can be accessed from a smartphone or computer. We are currently designing a prototype retinal camera rig which can house the smartphone and stabilize a patient for efficient imaging by a technician. The retinal camera rig prototype is built with PVC for the frame. We have produced CAD designs\footnote{The CAD designs are available at \href{https://github.com/Drishti-BD/Drishti-CAD}{https://github.com/Drishti-BD/Drishti-CAD}} for the remaining components. Ultimately, this would create a cost-effective and accurate solution for the widespread lack of access to DR screening in Bangladesh. 

\section{Acknowledgements}

We would like to thank Professor Malay Mridha from BRAC University James P. Grant School of Public Health for providing the images used in this validation study. We used secondary data from the project titled “understanding the pattern and determinants of health of South Asian people-South Asia Biobank”. This project was funded by the National Institute for Health Research (NIHR) (16/136/68) using UK aid from the UK Government to support global health research, and by Wellcome Trust (212945/Z/18/Z). The views expressed in this publication are those of us, the authors, and not necessarily those of the NIHR or the UK Department of Health and Social Care. Lastly, we would like to acknowledge Dr. Sabrina Rahmatullah for providing 43 images from the Bangladesh Eye Hospital and diagnosing them herself. 

\bibliographystyle{unsrtnat}
\bibliography{references}

\end{document}